\documentstyle[emulateapj,apjfonts,epsf]{article}

\slugcomment{Accepted for publication by The Astrophysical Journal
Letters, December 7, 2000}

\begin{document}

\title{Discovery of Five Binary Radio Pulsars}

\author{F.~Camilo,\altaffilmark{1,2}
  A.~G.~Lyne,\altaffilmark{2}
  R.~N.~Manchester,\altaffilmark{3}
  J.~F.~Bell,\altaffilmark{3}
  I.~H.~Stairs,\altaffilmark{2,4}
  N.~D'Amico,\altaffilmark{5,6}
  V.~M.~Kaspi,\altaffilmark{7,8}
  A.~Possenti,\altaffilmark{5}
  F.~Crawford,\altaffilmark{8}
  and N.~P.~F.~McKay\altaffilmark{2}}

\altaffiltext{1}{Columbia Astrophysics Laboratory, Columbia
  University, 550 W. 120th Street, New York, NY~10027;
  fernando@astro.columbia.edu}
\altaffiltext{2}{University of Manchester, Jodrell Bank Observatory,
  Macclesfield, Cheshire, SK11~9DL, UK}
\altaffiltext{3}{Australia Telescope National Facility, CSIRO,
  P.O.~Box~76, Epping, NSW~1710, Australia}
\altaffiltext{4}{National Radio Astronomy Observatory, P.O.~Box~2,
  Green Bank, WV~24944}
\altaffiltext{5}{Osservatorio Astronomico di Bologna, via
  Ranzani 1, 40127~Bologna, Italy}
\altaffiltext{6}{Istituto di Radioastronomia del CNR, via Gobetti
  101, 40129~Bologna, Italy}
\altaffiltext{7}{Department of Physics, Rutherford Physics Building,
  McGill University, 3600 University Street, Montreal, Quebec, H3A~2T8,
  Canada}
\altaffiltext{8}{Department of Physics and Center for Space
  Research, Massachusetts Institute of Technology, Cambridge, MA~02139}

\begin{abstract}
We report on five binary pulsars discovered in the Parkes multibeam
Galactic plane survey.  All of the pulsars are old, with characteristic
ages 1--11$\times10^9$\,yr, and have relatively small inferred magnetic
fields, 5--90$\times10^8$\,G.  The orbital periods range from 1.3 to 15
days.  As a group these objects differ from the usual low-mass binary
pulsars (LMBPs): their spin periods of 9--88\,ms are relatively long;
their companion masses, 0.2--1.1\,M$_\odot$, are, in at least some
cases, suggestive of CO or more massive white dwarfs; and some of the
orbital eccentricities, $10^{-5} \la e \la 0.002$, are unexpectedly
large.  We argue that these observed characteristics reflect binary
evolution that is significantly different from that of LMBPs.  We also
note that intermediate-mass binary pulsars apparently have a smaller
scale-height than LMBPs.

\end{abstract}

\keywords{Binaries: general --- pulsars: individual (PSR~J1232$-$6501,
PSR~J1435$-$6100, PSR~J1454$-$5846, PSR~J1810$-$2005, PSR~J1904+0412)}

\section{Introduction}\label{sec:intro}

Most of the $\sim 40$ binary pulsars known in the disk of the Galaxy
are millisecond pulsars with weak magnetic fields ($B \sim 10^8$\,G),
spin periods $2<P<15$\,ms, and in nearly circular orbits with
companions of mass $0.15 \la m_2 \la 0.4\,\mbox{M}_\odot$, presumably
He white dwarfs (WDs), some of which have been detected optically.
These are the low-mass binary pulsars (LMBPs), and their formation
mechanism is well understood.  After a neutron star spins down to long
periods and a low-mass companion evolves off the main sequence, a long
phase of stable mass-transfer ensues, during which the system may be
detectable as a low-mass X-ray binary (LMXB; see \cite{ver93} for a
review).  Eventually the orbit is circularized (\cite{phi92b}), the
pulsar spins up, its magnetic field is somehow quenched (e.g.,
\cite{rom90}), and a long-lived `recycled' radio millisecond pulsar
emerges.  Despite some uncertainties, it appears that the birth-rates
of LMXBs and LMBPs are comparable (\cite{lor00}), and this evolutionary
model successfully accounts for many properties of LMBPs.  However, it
should be noted that 20\% of millisecond pulsars are isolated, and it
is not clear how they have lost their presumed past companions.

A small but growing group of binary pulsars consists of objects with
$15<P<200$\,ms, intermediate-mass companions ($m_2 \ga
0.5\,\mbox{M}_\odot$, likely CO or heavier WDs), and orbital
eccentricities in some cases much larger than their LMBP counterparts.
These are the intermediate-mass binary pulsars (IMBPs), and it is not
entirely clear how they fit into the evolutionary scheme outlined
above.  It has been suggested that such systems undergo a period of
unstable mass-transfer and common-envelope (CE) evolution
(\cite{vdh94}).  IMBPs may have more in common with the evolution of
high-mass systems that spend part of their lives as high-mass X-ray
binaries (HMXBs) and are progenitors to eccentric-orbit double-neutron
star binaries, with the difference that they were not sufficiently
massive for a second supernova to have occurred.

The vast majority of millisecond pulsars known in the Galactic disk is
located within 2\,kpc of the Sun.  This is due to the loss of
sensitivity of most surveys at larger distances, particularly along the
Galactic plane.  To probe the Galaxy-wide distribution of LMBPs and to
learn more about rare species of pulsars it is therefore desirable to
search the distant Galactic plane with improved sensitivity.

The Parkes multibeam survey (\cite{lcm+00}; \cite{mlc+01}) covers a
region of the inner Galactic plane ($|b|<5\arcdeg$,
$-100\arcdeg<l<50\arcdeg$) with sensitivity far surpassing that of
previous pulsar surveys.  The main aim of the survey is to find young
and distant pulsars, but it retains good sensitivity to fast-spinning
pulsars.  A radio frequency of 1374\,MHz is used, reducing deleterious
propagation effects that affect the detectability of distant pulsars at
low latitudes.  So far, the survey has discovered more than 500 pulsars
(\cite{clm+00}; \cite{mlc+00}), including binary (\cite{lcm+00};
\cite{klm+00a}) and young (\cite{ckl+00}) pulsars.

In this {\em Letter\/} we report the discovery of five short-period
pulsars in binary systems.  They contribute significantly to our
understanding of binary pulsar evolution and demographics.

\section{Observations and Results}\label{sec:obs}

The survey uses the 13-beam receiver system at the 64-m Parkes
telescope in NSW, Australia.  Radio noise at a central frequency of
1374\,MHz and spanning 288\,MHz in bandwidth is filtered in a
$96\times3$-MHz filter bank spectrometer in each of two linear
polarizations, in observations lasting 35\,min.  Signals from
complementary polarizations are added, and the 96 voltages for each
beam are sampled every $250\,\mu$s, digitized, and written to magnetic
tape for off-line analysis.  The data are then searched for periodic
and dispersed signals using standard techniques (e.g., \cite{mld+96}).

Pulsars J1435$-$6100, J1810$-$2005, J1454$-$5846, J1232$-$6501, and
J1904+0412 were first detected in data collected on 1997 May 26, August
26, 1998 January 22, 24, and August 12, respectively.  Following
confirming observations, PSR~J1810$-$2005 has been monitored in a
series of timing observations with the 76-m Lovell telescope at Jodrell
Bank, UK, while the remaining pulsars have been observed at Parkes.

At Parkes we record data from the central beam in a manner otherwise
identical to the survey observations, while tracking each pulsar for
about 15\,min on each observing day, with the exception that since
MJD~51630 we have observed PSR~J1435$-$6100 with a $512\times0.5$-MHz
filter bank and a sampling interval of $125\,\mu$s at a central
frequency of 1390\,MHz.  Data are collected on a few days about every
two months, coinciding with epochs during which survey observations are
in progress.  Also, PSR~J1904+0412 was observed on a monthly basis with
the 305-m Arecibo telescope, from 1999 October through 2000 July, using
the Penn State Pulsar Machine, a $128\times0.0625$\,MHz filter bank
with $80\,\mu$s sampling at a central frequency of 1400\,MHz.  The
data, time-tagged with the start-time of the observations, are
de-dispersed and folded at the predicted topocentric pulsar period,
forming pulse profiles; and pulse times-of-arrival (TOAs) are measured
by cross-correlating these profiles with high signal-to-noise ratio
(S/N) templates (Fig.~\ref{fig:profs}), created from the addition of
many profiles.  Similar procedures are used at Jodrell Bank, with the
difference that the data are de-dispersed and folded on-line; also,
$32\times3$-MHz filter banks were used until MJD~51400, and
$64\times1$-MHz filter banks have been used since.

\medskip
\epsfxsize=8truecm
\epsfbox{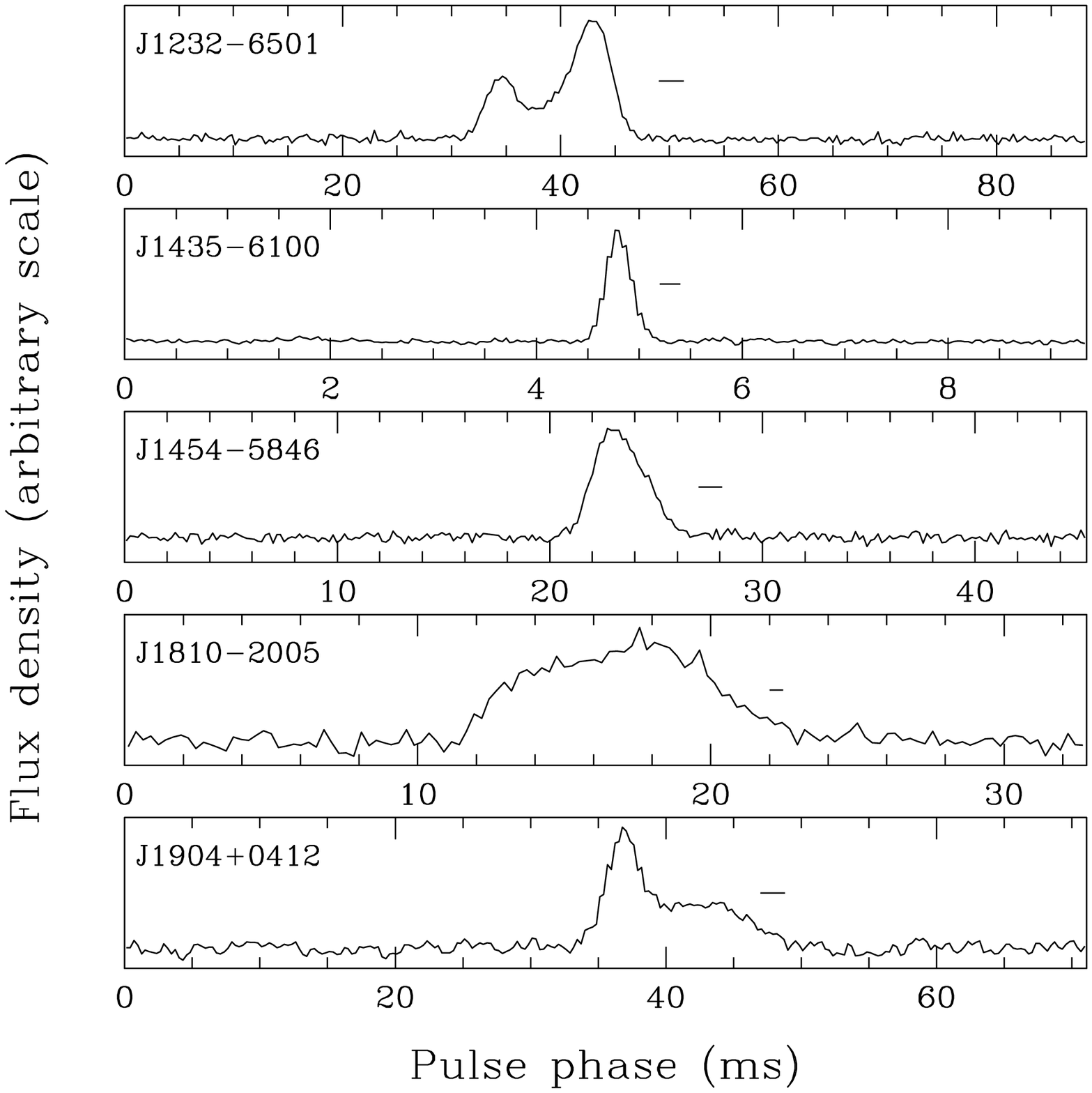}
\figcaption[profs.eps]{\label{fig:profs} Integrated pulse profiles for
five pulsars at a frequency of 1374\,MHz.  The time resolution of each
profile is indicated by a horizontal bar.  The profiles for
PSRs~J1232$-$6501, J1454$-$5846, and J1904+0412 are the template
profiles used to obtain TOAs.  That for J1435$-$6100, at 1390\,MHz, is
the template for the high-resolution data obtained since MJD~51630,
while that for J1810$-$2005 has time resolution a factor of six better
than the template and most of the data used to obtain its timing
solution. }

\bigskip

We then use the {\sc tempo} timing software\footnote{See
http://pulsar.princeton.edu/tempo.} to determine celestial coordinates,
spin, and orbital parameters for the pulsars.  This is done by first
converting the measured TOAs to the barycenter using initial estimates
of pulsar parameters and the DE200 solar-system ephemeris
(\cite{sta82}), and by minimizing timing residuals with respect to the
model parameters.  The parameters thus obtained are listed in
Table~\ref{tab:parms}, and the corresponding residuals are displayed in
Figure~\ref{fig:res1} as a function of date.

\medskip
\epsfxsize=8truecm
\epsfbox{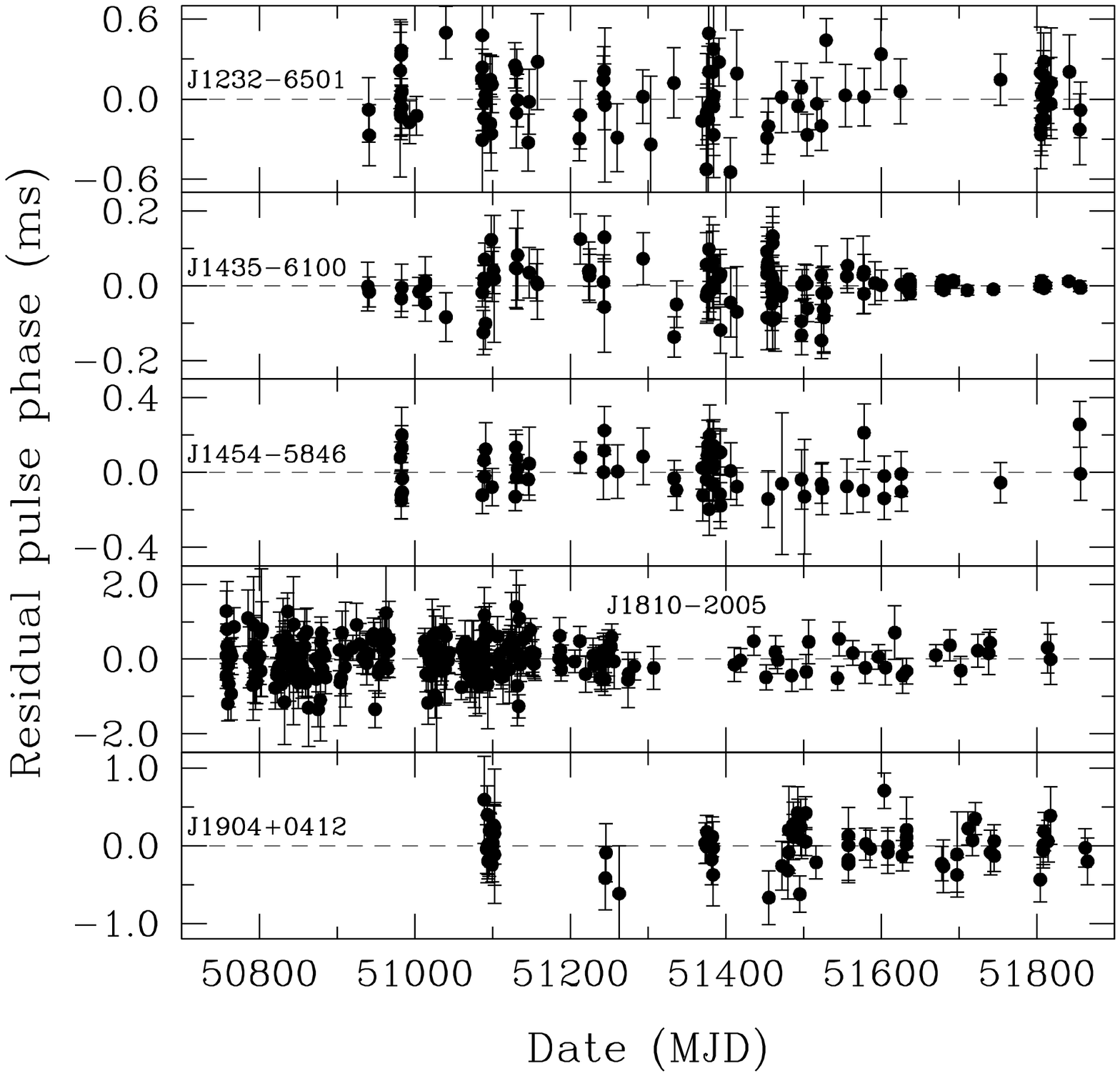}
\figcaption[res1.eps]{\label{fig:res1} Post-fit timing residuals as a
function of date for the five binary pulsars.  All orbits have been
well sampled, and residuals as a function of binary phase are
featureless.  }

\bigskip

The average flux densities listed in Table~\ref{tab:parms} were
estimated by converting the observed S/N to a scale calibrated using
stable flux densities known for a group of high-dispersion measure (DM)
pulsars.  See Manchester et al.~(2001)\nocite{mlc+01} for further
details of search and timing procedures.

\section{Discussion}\label{sec:disc}

\subsection{Evolution of the new systems}\label{sec:evol}

All five of the newly discovered pulsars have low inferred magnetic
fields ($B<10^{10}$\,G; Table~\ref{tab:parms}) when compared with the
vast majority of known pulsars (see Fig.~\ref{fig:ppdot}), and all are
in circular binary systems.  These characteristics indicate that all of
the pulsars have interacted with their companions in the past, and have
been recycled to some extent.  However, their periods and period
derivatives (and hence $B$) are larger than those of most millisecond
pulsars, as indicated in the $P$--$\dot P$ diagram of
Figure~\ref{fig:ppdot}: the spin parameters of PSR~J1435$-$6100 place
it marginally within the group of LMBPs at the lower left of the
diagram, while those for the remaining four pulsars place them squarely
amidst the IMBPs and double-neutron star systems.

\medskip
\epsfxsize=8truecm
\epsfbox{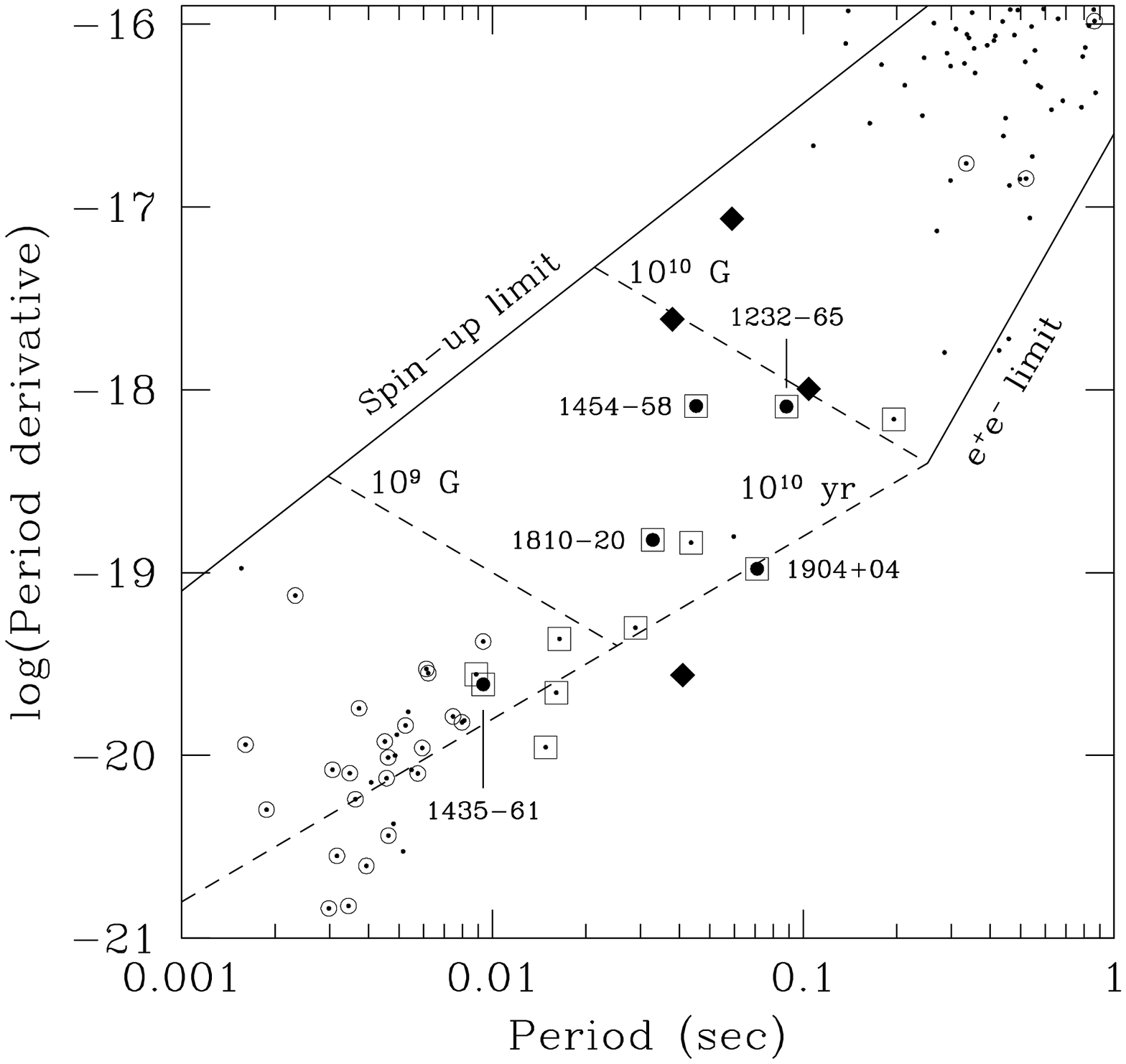}
\figcaption[ppdot.eps]{\label{fig:ppdot} Observed period derivative
vs.  period for the subset of pulsars in the Galactic disk with small
period derivatives (more than 1000 known pulsars lie above $\dot P >
10^{-16}$).  Dots denote isolated pulsars, circles indicate LMBPs,
squares represent IMBPs, and diamonds depict high-eccentricity,
double-neutron star binaries (see text).  Large dots represent the
pulsars presented in this paper, labeled by their partial names.  Two
lines of constant inferred magnetic field strength and a line of
characteristic age equal to $10^{10}$\,yr are indicated.  Pulsars spun
up via mass accretion must reside to the right of the spin-up limit
(see \cite{acw99} for a discussion).  }

\bigskip

Using the companion masses to attempt a classification of the new
systems yields results which are mostly inconsistent with those derived
from the spin parameters:  PSR~J1435$-$6100 has $m_2 \sim
1.1$\,M$_\odot$, decidedly not compatible with a LMBP; of the remaining
four systems only PSR~J1454$-$5846 ($m_2 \sim 1.1$\,M$_\odot$) appears
to be an IMBP, while the other three have $0.2 \la m_2 \la
0.3$\,M$_\odot$\footnote{These estimates of $m_2$ assume $m_1 =
1.35$\,M$_\odot$ (\cite{tc99}) and $i=60\arcdeg$; it is unlikely that
more than one of the three systems is sufficiently face-on so as to
have a $\ga 0.4$\,M$_\odot$ CO WD companion.} --- on this basis they
should be classified as LMBPs, but their periods and magnetic fields
are significantly larger than those of any LMBPs with remotely
comparable binary periods.

One further piece of useful information is provided by the orbital
eccentricities.  Phinney~(1992)\nocite{phi92b} derived a relationship
between eccentricity and binary period for LMBPs with $P_b \ga 2$\,d
that is remarkably consistent with observations.  One key ingredient of
the theory is that mass transfer to the neutron star via Roche-lobe
overflow be stable over the giant phase of evolution of the companion
star.  The relationship need therefore not hold for IMBPs
(\cite{pk94}), and for three of the five IMBPs with measured
eccentricities identified so far (\cite{cnst96}; \cite{ts99a};
\cite{eb00}) it does not (see Fig.~\ref{fig:epb}).

\subsubsection{Low-mass systems: non-standard evolution?}
\label{sec:low-mass}

Tauris \& Savonije~(1999)\nocite{ts99a} considered the detailed
non-conservative evolution of close binary systems with 1--2\,M$_\odot$
donor stars and accreting neutron stars, refining the well-known
correlation between $P_b$ and $m_2$ for LMBPs (\cite{jrl87}).  The
three new low-mass systems (PSRs~J1232$-$6501, J1810$-$2005, and
J1904+0412) follow this relation, considering the uncertainties in
$m_2$.

\medskip
\epsfxsize=8truecm
\epsfbox{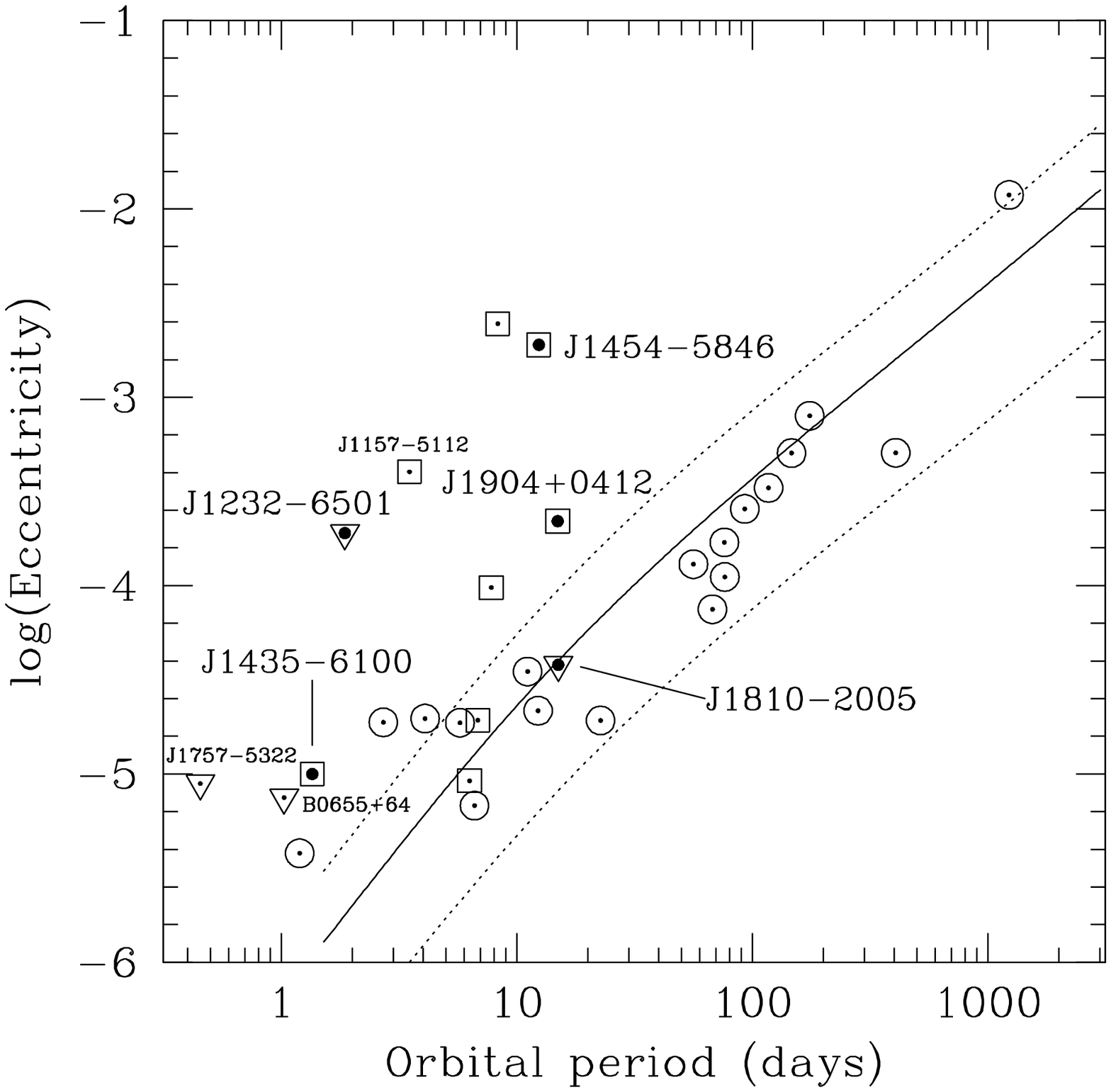}
\figcaption[epb.eps]{\label{fig:epb} Orbital eccentricity vs. period of
binary pulsars in the disk of the Galaxy with measured eccentricities
$e<0.1$.  Symbols are as in Figure~\protect\ref{fig:ppdot}, with
triangles denoting upper limits.  Three IMBPs mentioned in the text are
identified in small type.  The dotted lines should contain 95\% of the
eccentricities of the LMBPs (circles), according to the model of
Phinney \& Kulkarni~(1994).  \protect\nocite{pk94} }

\bigskip

Tauris, van~den~Heuvel, \& Savonije~(2000)\nocite{tvs00} then extended
this work to intermediate-mass (2--6\,M$_\odot$) donor stars.
Remarkably, they find that for a certain range of initial orbital
periods, such close binaries can survive periods of super-Eddington
mass transfer on sub-thermal (few Myr) timescales without experiencing
a CE phase.  Depending on initial donor mass and orbital period,
low-mass systems like the three we have discovered may result (see
their Figs.~2 and 4).

How shall we choose between these two alternative scenarios (low-
versus intermediate-mass original companions)?  Despite their present
low-mass companions, the newly discovered systems are unlikely to be
standard LMBPs, as already noted, because of their relatively large $P$
and $B$ (PSR~J1904+0412 also has too large an eccentricity;
Fig.~\ref{fig:epb}).  The intermediate-mass donor branch of evolution
is therefore more suitable to explain the new systems:  the
intermediate-mass systems tend to have shorter and less stable periods
of accretion, often at much higher rates, leading to a natural
explanation for the larger $P$, $B$, and (in at least some cases)
eccentricities.  With this evolutionary path, there is no need for a
long-lived X-ray accretion phase.  These systems might therefore not be
descendants of standard LMXBs, and should be accounted for separately
in birth-rate calculations.  What the X-ray progenitors of such systems
look like is of course an interesting and unresolved question.

\subsubsection{High-mass systems: common-envelope and a puzzle} 
\label{sec:high-mass}

As is clear from Figure~\ref{fig:epb}, the eccentricity of
PSR~J1454$-$5846 is much higher than predicted by the convective
fluctuation--dissipation theory of Phinney~(1992).\nocite{phi92b}  The
pulsar therefore has $P$, $B$, and eccentricity larger than expected
for LMBPs, and $m_2 \sim 1.1$\,M$_\odot$.  We thus confidently classify
it as an IMBP with a presumed O-Ne-Mg WD companion.  It is likely to
have undergone CE evolution and spiraled-in to its present $P_b =
12.4$\,d from an initial period of several hundred days, with a
companion of original mass 5--7\,M$_\odot$ (\cite{dt00};
\cite{tvs00}).  Edwards \& Bailes~(2000)\nocite{eb00} recently reported
the discovery of PSR~J1157$-$5112, a system broadly comparable to
PSR~J1454$-$5846, albeit with $P_b = 3.5$\,d and possibly a somewhat
larger companion mass.

The pulsar J1435$-$6100 is likely to have a massive ($m_2 \ga
1$\,M$_\odot$) O-Ne-Mg WD companion, like PSR~J1454$-$5846.  It must
have started with a very large orbital period so as not to coalesce
during the CE/spiral-in phase, and ended with $P_b = 1.35$\,d, much
smaller than $P_b = 12.4$\,d for PSR~J1454$-$5846.  A difficulty with
understanding PSR~J1435$-$6100 lies in its spin parameters:  they are
closer to those of LMBPs than IMBPs (see Fig.~\ref{fig:ppdot}).  In
other words, despite a presumed short-lived ($\sim 10^4$\,yr) mass
transfer phase in a CE (and hence very little accretion), the pulsar's
magnetic field was somehow quenched to a very low value
($5\times10^8$\,G), while it was spun up to a fast initial rate ($P_i
\la 9$\,ms).  Compare its parameters with those of the IMBP~B0655+64:
$P_b = 1.3$ vs. 1.0\,d; $m_2 \sim 1.1$ vs.  0.8\,M$_\odot$; both with
similar eccentricities, and likely products of CE evolution.  While the
orbital parameters are thus fairly similar, the spin parameters are the
most different within IMBPs: both $B$ and the present-day period of
PSR~B0655+64 are 23 times larger than those of PSR~J1435$-$6100.  The
recently discovered PSR~J1757$-$5322 (\cite{eb00}) has spin parameters
virtually identical to those of PSR~J1435$-$6100 (Fig.~\ref{fig:ppdot})
and orbital parameters also similar to those of PSR~B0655+64.  The
reason behind such contrasting sets of parameters between
PSRs~J1435$-$6100/J1757$-$5322 and B0655+64 is a puzzle.

\subsection{The scale-height of IMBPs and LMBPs}\label{sec:z}

The preceding discussion suggests that classifying pulsars by
present-day companion mass alone may not be particularly useful.  We
therefore define IMBPs as objects that once had intermediate-mass donor
stars.  While this is a model-dependent definition, operationally it
applies to pulsar systems with $10 \la P \la 200$\,ms and $e \la
0.01$.  Among these systems most, but not all (e.g., PSR~J1232$-$6501),
have $m_2 \ga 0.4$\,M$_\odot$, and $0.5 \la P_b \la 15$\,d.

It is notable that seven of the 12 presently known IMBPs (squares in
Fig.~\ref{fig:ppdot}) have been discovered in recent low- or
intermediate-latitude surveys (this {\em Letter\/} and \cite{eb00}).
This is despite the greater effective volume searched with at least
comparable sensitivity to pulsars with $P \ga 10$\,ms in some `all-sky'
surveys (e.g., \cite{cnt96}; \cite{lml+98}).  We now address this
curiosity.

The median distance perpendicular from the Galactic plane for 23 known
LMBPs is $|z|_{\rm m} = 0.4$\,kpc, and three systems have
$|z|>1.8$\,kpc (\cite{cam99}).  For the group of 12 IMBPs, $|z|_{\rm m}
= 0.2$\,kpc and the largest distance is $|z| = 0.5$\,kpc (\cite{cam99};
\cite{eb00}; this {\em Letter\/}).  Despite selection effects affecting
these determinations for both populations, it appears that IMBPs have a
smaller scale-height than LMBPs.  The maximum perpendicular distance
that a pulsar born near the plane can reach is approximately
proportional to the square of its initial perpendicular velocity.  A
scale-height for IMBPs that may be a factor of 2--4 smaller than for
LMBPs requires a velocity for IMBPs a factor of $\la 2$ smaller than
for LMBPs.  This is plausible, considering that a typical LMBP
progenitor is a $1 + 1.3$\,M$_\odot$ system while an IMBP may descend
from a $4 + 1.3$\,M$_\odot$ system.  In summary, the recent flurry of
IMBP discoveries may be due simply to the fact that recent efforts are
surveying with significant sensitivity where IMBPs tend to reside ---
along the Galactic plane.  Similar distributions apply to X-ray
binaries:  HMXBs have smaller average velocity and scale-height than
LMXBs (\cite{vpm95}).

The newly discovered IMBPs are distant objects ($3 \la d \la 10$\,kpc),
and were detected because they are relatively luminous pulsars ($2 \la
L_{1400} \la 30$\,mJy\,kpc$^2$; Table~\ref{tab:parms})\footnote{For
typical spectral indices, these would correspond to luminosities at
400\,MHz about 10 times greater, $20 \la L_{400} \la
300$\,mJy\,kpc$^2$.  For comparison, a tabulation of 21 millisecond
pulsars at $d \la 1.5$\,kpc has median $L_{400} = 10$\,mJy\,kpc$^2$ and
only one with $L_{400} > 20$\,mJy\,kpc$^2$ (\cite{lor00}).}.  Therefore
they need not contribute greatly to the overall population of binary
pulsars in the Galaxy.   However, in order to determine conclusively
the scale-height of IMBPs, and their incidence among binary pulsars, it
is necessary to perform careful modeling of the recent high-frequency
surveys, and to measure proper motions where possible.

\acknowledgements

We are grateful to T.~Tauris for many enlightening discussions, and
thank R.~Bhat, P.~Freire, G.~Hobbs, M.~Kramer, D.~Kubik, D.~Lorimer,
M.~McLaughlin, and D.~Morris for assistance with observations.  The
Parkes radio telescope is part of the Australia Telescope which is
funded by the Commonwealth of Australia for operation as a National
Facility managed by CSIRO.  Arecibo Observatory is part of the National
Astronomy and Ionosphere Center, which is operated by Cornell
University under a cooperative agreement with the US National Science
Foundation (NSF).  F.~Camilo is supported by NASA grant NAG~5-3229.
IHS received support from a Natural Sciences and Engineering Research
Council of Canada (NSERC) postdoctoral fellowship.  VMK was supported
in part by an Alfred P. Sloan Research Fellowship, NSF Career Award
(AST-9875897), and NSERC grant RGPIN~228738-00.

\begin{deluxetable}{llllll}
\tablewidth{0pt}
\footnotesize
\tablecaption{\label{tab:parms}Parameters for five binary pulsars}
\tablecolumns{6}
\tablehead{
\colhead{}                    &
\colhead{PSR~J1232$-$6501}    &
\colhead{PSR~J1435$-$6100}    &
\colhead{PSR~J1454$-$5846}    &
\colhead{PSR~J1810$-$2005}    &
\colhead{PSR~J1904$+$0412}    }
\startdata
R. A. (J2000)\dotfill                   & 12~32~17.840(5) & 14~35~20.2765(4)
                                        & 14~54~10.908(2) & 18~10~58.988(2)
                                        & 19~04~31.382(4)                    \nl
Decl. (J2000)\dotfill                   & $-65$~01~03.33(4) & $-61$~00~57.956(6)
                                        & $-58$~46~34.74(3) & $-20$~05~08.3(6)
                                        & $+04$~12~05.9(1)                   \nl
Period, $P$ (ms)\dotfill            & 88.2819082341(3) & 9.347972210248(6)
                                    & 45.24877299802(9) & 32.82224432571(9)
                                    & 71.0948973807(3)                       \nl
Period derivative, $\dot P$\dotfill & $8.1(2)\times 10^{-19}$
                                    & $2.45(4)\times 10^{-20}$
                                    & $8.16(7)\times 10^{-19}$ 
                                    & $1.51(7)\times 10^{-19}$
                                    & $1.1(3)\times 10^{-19}$                \nl
Epoch (MJD)\dotfill                 & 51270.0 & 51270.0 & 51300.0 & 51200.0 
                                    & 51450.0                                \nl
Orbital period, $P_b$ (days)\dotfill                & 1.86327241(8) 
                                                    & 1.354885217(2)
                                                    & 12.4230655(2)
                                                    & 15.0120197(9)
                                                    & 14.934263(2)           \nl
Projected semi-major axis, $x$ (l-s)\dotfill                 & 1.61402(6) 
                                                             & 6.184023(4)
                                                             & 26.52890(4)
                                                             & 11.97791(8)
                                                             & 9.6348(1)     \nl
Eccentricity, $e$\dotfill                           & 0.00011(8)
                                                    & 0.000010(2)
                                                    & 0.001898(3)
                                                    & 0.000025(13)
                                                    & 0.00022(2)             \nl
Time of ascending node, $T_{\rm asc}$ (MJD)\tablenotemark{a}\dotfill 
                                                    & 51269.98417(2)
                                                    & 51270.6084449(5)
                                                    & 51303.833(4)
                                                    & 51198.92979(2)
                                                    & 51449.45(25)           \nl
Longitude of periastron, $\omega$ (deg)\dotfill     & 129(45)
                                                    & 10(6)
                                                    & 310.1(1)
                                                    & 159(30)
                                                    & 350(6)                 \nl
Span of timing data (MJD)\dotfill & 50940--51856 & 50939--51856 & 50981--51856 
                                  & 50757--51817 & 51089--51865              \nl
Weighted rms timing residual ($\mu$s)\dotfill  & 200 & 14 & 100 & 430 & 240  \nl
Dispersion measure, DM (cm$^{-3}$\,pc)\dotfill & 239.4(5) 
                                               & 113.7(6) 
                                               & 116.0(2) 
                                               & 240.2(3)
                                               & 185.9(7)                    \nl
Flux density at 1400\,MHz, $S_{1400}$ (mJy)\dotfill 
                                               & 0.3 & 0.2 & 0.2 & 1.1 & 0.3 \nl
\cutinhead{Derived parameters\tablenotemark{b}}
Galactic longitude, $l$ ($\deg$)\dotfill 
                            & 300.9   & 315.2   & 318.3  & 10.5    & 38.0    \nl
Galactic latitude, $b$ ($\deg$)\dotfill  
                            & $-2.2$  & $-0.6$  &  0.4   & $-0.6$  & $-1.1$  \nl
Surface magnetic field strength, $B$ ($10^8$\,G)\dotfill 
                                                    & 90 & 5 & 60  & 20 & 30 \nl
Characteristic age, $\tau_{\rm c}$ ($10^9$\,yr)\dotfill 
                                                    &  2 & 6 & 0.9 &  3 & 11 \nl
Mass function, $f_1$ (M$_{\odot}$)\dotfill
                            & 0.0013  & 0.1383  & 0.1299 & 0.0082  & 0.0043  \nl
Companion mass, $m_2$ (M$_{\odot}$)\dotfill
                            & $>0.14$ & $>0.90$ &$>0.87$ & $>0.28$ & $>0.22$ \nl
Distance, $d$ (kpc)\dotfill & 10      & 3.3     & 3.3    & 4.0     & 4.0     \nl
Distance from Galactic plane, $|z|$ (kpc)\dotfill 
                            & 0.4     & 0.03    & 0.02   & 0.04    & 0.08    \nl
Radio luminosity, $L_{1400}$ (mJy\,kpc$^2$)\dotfill
                            & 30      & 2       & 2      & 18      & 5       \nl
\enddata

\tablecomments{Units of right ascension are hours, minutes, and
seconds, and units of declination are degrees, arcminutes, and
arcseconds.  Figures in parentheses are twice the nominal {\sc tempo}
uncertainties in the least-significant digits quoted, obtained after
scaling TOA uncertainties to ensure $\chi ^2_{\nu} = 1$. }

\tablenotetext{a}{Due to the large covariance between $\omega$ and time
of periastron $(T_0)$ in standard {\sc tempo} fits for pulsars with $e
\ll 1$, the solutions for PSRs~J1232$-$6501, J1435$-$6100, and
J1810$-$2005 were obtained using the ELL1 model, where $T_{\rm asc}
(\omega \equiv 0)$ and $(e \cos \omega, e \sin \omega)$ are fit instead
(\cite{lcw+01}).  In these cases $e$ and $\omega$ (as well as $T_0$)
can be derived.  For the other two pulsars we used the standard (BT)
binary model that fits for $e$, $\omega$, and $T_0$ --- which is listed
instead of $T_{\rm asc}$. }

\tablenotetext{b}{The following formulae are used to derive some
parameters: $B=3.2\times10^{19} (P\dot P)^{1/2}$\,G; $\tau_{\rm c} =
P/(2\dot P)$; and $f_1 = x^3 (2\pi/P_b)^2 T_\odot^{-1} = (m_2\sin
i)^3/(m_1+m_2)^2$, where $T_\odot \equiv G M_\odot/c^3 = 4.925\,\mu$s,
$m_1$ and $m_2$ are the pulsar and companion masses, respectively, and
$i$ is the orbital inclination angle.  $m_2$ is obtained from the mass
function, with $m_1 = 1.35$\,M$_{\odot}$ (\cite{tc99}) and
$i<90\arcdeg$.  The distances are calculated from the DMs with the
Taylor \& Cordes~(1993)\nocite{tc93} free-electron distribution model;
$|z| = d \sin |b|$; and $L_{1400} = S_{1400} d^2$. }

\end{deluxetable}


\begin{thebibliography}{}

\bibitem[Arzoumanian, Cordes, \& Wasserman 1999]{acw99}
Arzoumanian, Z., Cordes, J.~M., \& Wasserman, I. 1999, ApJ, 520, 696

\bibitem[{Camilo} 1999]{cam99}
{Camilo}, F. 1999, in { Pulsar Timing, General Relativity, and the Internal
  Structure of Neutron Stars}, ed.\ Z. Arzoumanian, F. {van der Hooft}, \&
  E.~P.~J. {van den Heuvel}, (Amsterdam: North Holland), 115

\bibitem[Camilo {et al.}  2000a]{ckl+00}
Camilo, F., Kaspi, V.~M., Lyne, A.~G., Manchester, R.~N., Bell, J.~F., D'Amico,
  N., McKay, N. P.~F., \& Crawford, F. 2000a, ApJ, 541, 367

\bibitem[Camilo {et al.}  2000b]{clm+00}
Camilo, F. {et al.}  2000b, in { Pulsar Astronomy --- 2000 and Beyond, {IAU}
  Colloquium 177}, ed.\ M. Kramer, N. Wex, \& R. Wielebinski, (San Francisco:
  Astronomical Society of the Pacific), 3

\bibitem[Camilo {et al.}  1996]{cnst96}
Camilo, F., Nice, D.~J., Shrauner, J.~A., \& Taylor, J.~H. 1996, ApJ, 469, 819

\bibitem[Camilo, Nice, \& Taylor 1996]{cnt96}
Camilo, F., Nice, D.~J., \& Taylor, J.~H. 1996, ApJ, 461, 812

\bibitem[Dewi \& Tauris 2000]{dt00}
Dewi, J. D.~M. \& Tauris, T.~M. 2000, A\&A, 360, 1043

\bibitem[Edwards \& Bailes 2000]{eb00}
Edwards, R.~T. \& Bailes, M. 2000, ApJ.
\newblock Submitted, astro-ph/0010599

\bibitem[{Joss}, {Rappaport}, \& {Lewis} 1987]{jrl87}
{Joss}, P.~C., {Rappaport}, S., \& {Lewis}, W. 1987, ApJ, 319, 180

\bibitem[Kaspi {et al.}  2000]{klm+00a}
Kaspi, V.~M. {et al.}  2000, ApJ, 543, 321

\bibitem[Lange {et al.}  2001]{lcw+01}
Lange, C., Camilo, F., Wex, N., Kramer, M., Backer, D.~C., Lyne, A.~G., \&
  Doroshenko, O. 2001, MNRAS.
\newblock Submitted

\bibitem[Lorimer 2000]{lor00}
Lorimer, D.~R. 2000, in { {The Neutron Star-Black Hole Connection} ({NATO ASI
  Series})}, ed.\ J. Ventura, (Dordrecht: Kluwer), astro-ph/9911519

\bibitem[Lyne {et al.}  2000]{lcm+00}
Lyne, A.~G. {et al.}  2000, MNRAS, 312, 698

\bibitem[Lyne {et al.}  1998]{lml+98}
Lyne, A.~G. {et al.}  1998, MNRAS, 295, 743

\bibitem[Manchester {et al.}  2001]{mlc+01}
Manchester, R.~N. {et al.}  2001, MNRAS.
\newblock Submitted

\bibitem[Manchester {et al.}  2000]{mlc+00}
Manchester, R.~N. {et al.}  2000, in { Pulsar Astronomy --- 2000 and Beyond,
  {IAU} Colloquium 177}, ed.\ M. Kramer, N. Wex, \& R. Wielebinski, (San
  Francisco: Astronomical Society of the Pacific), 49

\bibitem[Manchester {et al.}  1996]{mld+96}
Manchester, R.~N. {et al.}  1996, MNRAS, 279, 1235

\bibitem[Phinney 1992]{phi92b}
Phinney, E.~S. 1992, Philos. Trans. Roy. Soc. London A, 341, 39

\bibitem[Phinney \& Kulkarni 1994]{pk94}
Phinney, E.~S. \& Kulkarni, S.~R. 1994, ARAA, 32, 591

\bibitem[Romani 1990]{rom90}
Romani, R.~W. 1990, Nature, 347, 741

\bibitem[Standish 1982]{sta82}
Standish, E.~M. 1982, A\&A, 114, 297

\bibitem[Tauris \& Savonije 1999]{ts99a}
Tauris, T.~M. \& Savonije, G.~J. 1999, A\&A, 350, 928

\bibitem[Tauris, van~den Heuvel, \& Savonije 2000]{tvs00}
Tauris, T.~M., van~den Heuvel, E. P.~J., \& Savonije, G.~J. 2000, ApJ, 530, L93

\bibitem[Taylor \& Cordes 1993]{tc93}
Taylor, J.~H. \& Cordes, J.~M. 1993, ApJ, 411, 674

\bibitem[Thorsett \& Chakrabarty 1999]{tc99}
Thorsett, S.~E. \& Chakrabarty, D. 1999, ApJ, 512, 288

\bibitem[{van den Heuvel} 1994]{vdh94}
{van den Heuvel}, E. P.~J. 1994, A\&A, 291, L39

\bibitem[{van Paradijs} \& {McClintock} 1995]{vpm95}
{van Paradijs}, J. \& {McClintock}, J.~E. 1995, in { X-ray Binaries}, ed.\
  J.~van~Paradijs W.~H.~G.~Lewin \& E.~P.~J. van~den Heuvel, (Cambridge:
  Cambridge Univ. Press), 58

\bibitem[Verbunt 1993]{ver93}
Verbunt, F. 1993, ARAA, 31, 93

\end{thebibliography}
\end{document}